\documentclass[aps,pre,twocolumn,superscriptaddress,showpacs]{revtex4}
\usepackage{graphicx}
\begin{document}

%\preprint{}

\title{Two-phase flow in porous media: dynamical phase transition}


\author{Henning Arendt Knudsen}
\email[]{arendt@phys.ntnu.no}
\affiliation{Institute of Physics, University Duisburg-Essen, 
D-47048 Duisburg, Germany}
\affiliation{Department of Physics, Norwegian University of Science and 
Technology, NTNU, NO-7491 Trondheim, Norway}
\author{Alex Hansen}
\email[]{alexh@phys.ntnu.no}
%\thanks{}
\affiliation{Department of Physics, Norwegian University of Science and 
Technology, NTNU, NO-7491 Trondheim, Norway}


\date{June 11, 2003}

\begin{abstract}
Two-phase flow systems in porous media have complex dynamics. 
It is well established that a wide range of system parameters 
like viscosities and porosity as well as flow parameters such as 
pressure gradient and fluid saturation have strong impact on the dynamics. 
The transition from single-phase flow to two-phase flow is a dynamical 
phase transition. We discuss the order of the transition and investigate 
the phase diagram in parameter space, using a network simulator for two-phase 
flow. A semi-empirical theory for the location of the phase boundaries is 
provided.
\end{abstract}

%\keywords{}

\maketitle

\section{Introduction}
\label{sec:intro}
The field of two-phase flow in porous media is very rich on problems 
of a complex nature. These involve the study of properties on many 
different length scales and how to bridge the gap between the scales. 
Further, it is often possible to distinguish between the study of 
invasion processes and steady-state properties. The methods used in 
this field range from a set of experimental techniques, theoretical 
descriptions on different length scales to several numerical models or 
methods\cite{D92,S95}.

In the petroleum engineering tradition, the main source of experimental 
information about flow systems has for decades been displacement 
experiments on core samples\cite{TD96}. However, for increased theoretical 
or fundamental understanding, well controlled laboratory experiments have 
been the key. In particular, in the eighties, the experiments and simulations 
by Lenormand \emph{et al.} lead to the development of a phase diagram for 
drainage, i.e., the displacement of a wetting phase by a nonwetting 
phase\cite{LZS83,LTZ88}. Drainage was classified into the regimes stable 
displacement, viscous fingering, and capillary fingering. The regime 
boundaries depend on two dimensionless numbers, the capillary number Ca
and the viscosity ratio $M$.

Two-phase flow consists of more than pure displacement processes. The 
complexity of ganglion dynamics and the existence of different regimes 
of flow of bubbles and blobs has been investigated experimentally by the 
Payatakes group\cite{AP95a,AP95b,AP99}. They used etched glass networks 
for their studies, varying flow parameters within large ranges. By 
simultaneous injection of two fluids the experiments examine steady-state 
properties or close to steady-state properties.

We use a simulator for two-phase flow in porous media that is based on 
the Washburn equation\cite{W21}, see Sec.\ref{sec:model}. This line of 
modeling dates back to the mid eighties to Koplik \emph{et al.}\cite{KL85}. 
The work by the Payatakes group is a continuation of this 
tradition\cite{CP91,CP96}. Most of the work on two-phase flow simulation 
has been in the actual two-phase regime. That is to say that pure invasion 
processes like drainage were simulated until breakthrough of the nonwetting 
phase. The aforementioned regimes, the phase diagram, of drainage are well 
established and their properties much studied. When it comes to simulations 
for investigating steady-state properties, the majority of the work involves 
finding relative permeability curves for a wide range of parameters. Much 
effort has been put into making simulators that are specific for a given 
porous medium and fluid system. Further, other simulation techniques exist. 
On a even more detailed scale than our modeling is the work using lattice 
Boltzmann methods\cite{R90,FR95,LP01}.

We wish here to take one step back and take a broader look at these systems. 
Assuming a porous medium with two phases, one wetting and one nonwetting with 
respect to the medium, there are three possible states of flow: only the 
nonwetting phase flows, only the wetting phase flows, or both flow 
simultaneously. Depending on a large set of parameters the system will find 
itself within one of these situations. Changing system parameters within 
appropriate ranges causes the system to undergo what is called a dynamical 
phase transition. Using the language of thermodynamics and critical phenomena 
we provide in this paper a phase diagram for steady-state flow. As was the 
case for the phase diagram of Lenormand \emph{et al.} the capillary number 
and the viscosity ratio are the parameters of interest also in this primary 
study of the steady-state phase diagram.

We use a not too complex network simulator that can do real steady-state 
simulations. Nevertheless, the resulting phase diagram is complex. We believe 
that the structure of this diagram is of general interest. Phase boundaries 
are located by the simulations. Further, based on the simulations a 
semi-empirical theory for the location of the phase boundaries is given. 
We believe the overall structure of the phase diagram is universal for the 
two-phase flow system, whereas the exact positions of phase boundaries 
will vary from system to system.

%\subsection{}
%\subsubsection{}

\section{Model}
\label{sec:model}
In order to investigate question of interest regarding two-phase flow in 
porous media it is possible to apply methods that count as theoretical, 
experimental or numerical. Experimental methods provide quantitative `correct' 
results for the actual system investigated. However, results may differ 
from sample to sample or from system to system. By means of simulations 
the system properties can be controlled to a greater extent. This is 
highly advantageous when one wishes to study the effect of varying 
parameters of the system.

The results of the paper are based on a network simulator for 
immiscible two-phase flow. This line of modeling which is based on 
the Washburn equation\cite{W21} dates back to the work of several groups in 
the mid 1980's\cite{KL85,CP91,CP96}. Our model is a continuation of the 
model developed by Aker \emph{et al.}\cite{AMHB98,AMH98}. Although a 
thorough presentation can be found in \cite{KAH02}, we provide for 
clarity a brief r{\'e}sum{\'e} of the main aspects.

The porous media are represented by networks of tubes, forming square 
lattices in two dimensions (2D) and cubic lattices in 3D, both 
tilted $45^{\circ}\ $with respect to the imposed pressure gradient and 
thus to the overall direction of flow. We refer to the the lattice sites 
where four(2D) and six(3D) tubes meet as nodes. Volume in the model is 
contained in the tubes and not in the nodes, although effective node volumes 
are used in the modeling of transport through the nodes. For further details, 
see\cite{KAH02}. Randomness is incorporated by distorting the nodes 
randomly within a circle (2D) and a sphere (3D) around their respective 
lattice positions. This gives a distribution of tube lengths in the system. 
Further, the radii are drawn from a flat distribution so that the 
radius of a given tube is $r\in(0.1l,0.4l)$, where $l\ $is the length of that 
tube.

The model is filled with two fluid phases that flow within the system of 
tubes. The flow in each tube obeys the Washburn equation\cite{W21}, 
$q=-(\sigma k/\mu)(\Delta p-\sum{p_c})/l$. With respect to momentum transfer 
these tubes are cylindrical with cross-section area $\sigma$, length $l$, 
and radius $r$. The permeability is $k=r^2/8\ $which is known for 
Hagen-Poiseuille flow\cite{GGH92}. Further, $\mu\ $is the viscosity of 
the phase present in the tube. If both phases are present the volume 
average of their viscosities is used. The volumetric flow rate is denoted by 
$q\ $and the pressure difference between the ends of the tube by 
$\Delta p$. The summation is the sum over all capillary pressures 
$p_c\ $within the tube. With respect to capillary pressure the tubes are 
hour-glass shaped, meaning that a meniscus at position $x\in(0,l)\ $in the 
tube has capillary pressure $p_c=(2\gamma/r)[1-\cos{(2\pi x/l)}]$, where 
$\gamma\ $is the interfacial tension between the two phases. This is a 
modified version of the Young-Laplace law\cite{D92,AMHB98}.

Biperiodic (triperiodic in case of 3D) boundary conditions are used so 
that the flow is by construction steady flow. That is to say, the systems 
are closed so both phases retain their initial volume fractions, i.e., 
their saturations. These boundary conditions are in 2D equivalent 
to a flow restrained to be on the surface of a torus. The flow is driven 
by a globally applied pressure gradient. Usually invasion processes are 
driven by setting up a pressure difference between two borders, inlet and 
outlet. Since, by construction, the outlet is directly joined with the inlet 
in our system, we give instead a so-called global pressure drop when passing 
a line or cut through the system\cite{roux,KAH02}. This is equivalent to 
imposing constraints on the pressure gradient that is experienced throughout 
the network. Integration of the pressure gradient along an arbitrary 
closed path one lap around the system, making sure to pass the 
`inlet-outlet'-cut once, should add up to the same global pressure difference.

The system is forward integrated in time using the Euler scheme. For each 
time step the distribution of the phases leads to a recalculation of the 
effective viscosities in the tubes and the capillary pressures across the 
menisci, and thus the coefficients in the equations for the pressure field. 
When the pressure field is known, the flow field follows automatically and 
the integration step may be performed. Details are found in\cite{KAH02}.

\section{Phase diagrams}
\label{sec:diagram}
We present here phase diagrams for the dynamical properties of steady 
two-phase flow. The structure of the phase diagram is richer than one might 
naively believe. First we introduce the well-known concept of fractional 
flow (Eq.(\ref{eq:fflowdef})) and provide a basic description of the nature 
of the two-phase flow system. We determine phase boundaries in the parameter 
space of saturation $S_{\rm nw}$ (defined in Eq.(\ref{eq:satdef})), capillary 
number Ca (defined in Eq.(\ref{eq:cadef})), and viscosity ratio 
$M\ $(defined in Eq.(\ref{eq:mdef})). The definition of Ca is a crucial 
point since the definition is not unique. For the numerical exploration of 
the phase diagram one definition is used consistently in all simulation 
series: first at constant Ca and varying $S_{\rm nw}$, second at constant 
$S_{\rm nw}\ $and varying Ca. While Sec.\ \ref{sec:diagram} contains 
simulated diagrams, Sec.\ \ref{sec:theory} readdresses the definition of 
the capillary number and contains a semi-empirical theory for the location 
of the phase boundaries that are obtained in Sec.\ \ref{sec:diagram}. Upon 
presenting the method of finding the phase boundaries, the order of the 
transition is briefly discussed. However, the final discussion of the 
order is postponed to Sec.\ \ref{sec:ordertrans}.

\subsection{Order parameter}
\label{sec:order}

\begin{figure}[t!]
\includegraphics[width=6cm]{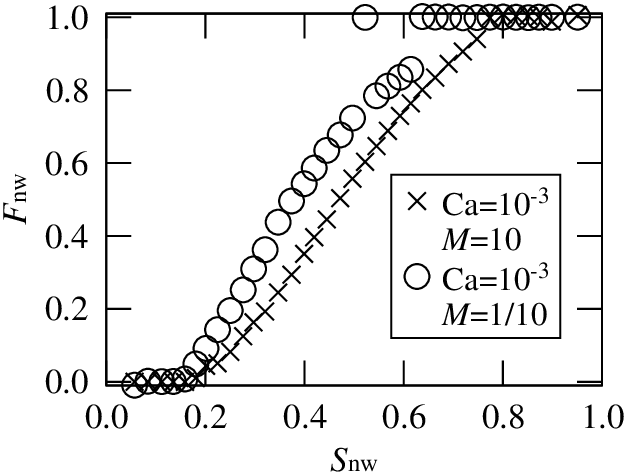}
\caption{The nonwetting fractional flow is shown as a function of 
nonwetting saturation for two sets of system parameters. Each data point 
is a result of simulation of the model system with the parameters 
in question until the system reaches steady state and the properties are 
measured. At low $S_{\rm nw}\ $only the wetting phase flows and at high 
$S_{\rm nw}\ $only the nonwetting phase flows. In-between there is a region 
of two-phase flow. We also observe that the transition of both curves 
from zero to nonzero $F_{\rm nw}\ $is continuous. The transition of 
$\times\ $from two-phase flow to pure nonwetting flow is continuous or 
weakly first order, while the same transition for $\circ\ $is first order. 
There seems to be a clear jump in the value of $F_{\rm nw}\ $ as well as 
indications of hysteresis.}
\label{fig:fofssample}
\end{figure}

In steady flow the volume fractions of the wetting and the nonwetting phase 
will not change in time. In our simulation model this requirement is 
fulfilled by having a closed system. The nonwetting saturation of the 
system is defined as
\begin{equation}
\label{eq:satdef}
S_{\rm nw} = \frac{V_{\rm nw}}{V_{\rm tot}},
\end{equation}
where $V_{\rm nw}\ $and $V_{\rm tot}\ $are the nonwetting and total volume, 
respectively. The wetting saturation $S_{\rm w}\ $is defined similarly. 
For each value of the saturation one can in the simulations fix the total 
flux $Q_{\rm tot}\ $in the system. In order to keep the total flux constant 
the imposed pressure gradient fluctuates in time around some typical mean 
value. This mean pressure is a function of saturation and total flux. 
Further, one can measure the flux of each of the phases in the system, 
$Q_{\rm nw}\ $and $Q_{\rm w}$, respectively. One defines the nonwetting 
fractional flow as
\begin{equation}
\label{eq:fflowdef}
F_{\rm nw} = \frac{Q_{\rm nw}}{Q_{\rm tot}},
\end{equation}
and likewise for the wetting fractional flow. A sample of nonwetting 
fractional flow as a function of nonwetting saturation is given in Fig.\ 
\ref{fig:fofssample}. The transport properties of the system depend on 
dimensionless groups of system parameters, namely the viscosity ratio and 
capillary number. We define the viscosity ratio as the nonwetting 
viscosity divided by the wetting viscosity;
\begin{equation}
\label{eq:mdef}
M=\frac{\mu_{\rm nw}}{\mu_{\rm w}}.
\end{equation}
Further, the capillary number is defined as
\begin{equation}
\label{eq:cadef}
{\rm Ca} = {\rm Ca_{dyn}} = \frac{Q_{\rm tot}\mu}{\gamma\Sigma},
\end{equation}
where $\gamma\ $is the interfacial tension between the two phases, $\Sigma\ 
$ is the cross-section area of the system, and $\mu\ $is the effective 
(weighted) viscosity of the system; 
$\mu=S_{\rm nw}\mu_{\rm nw}+S_{\rm w}\mu_{\rm w}$. 
Physically the capillary number measures the ratio between viscous and 
capillary forces in the system. Using the weighted viscosity is motivated by 
the fact that when both phases flow, both viscosities play a role in 
determining the relative strength of viscous and capillary forces. 
We use this \emph{dynamical} definition of Ca in this section, upon the 
exploration of the phase space. For convenience we leave out the subscript 
of ${\rm Ca_{dyn}}$, but we will use it in the subsequent discussion in Sec.\
\ref{sec:theory} for clarity.

As illustrated in Fig.\ \ref{fig:fofssample} fractional flow curves depend 
on $M\ $and Ca. This dependence is nontrivial. The curves exhibit three 
different regimes. For low nonwetting saturation only the wetting phase flows. 
Likewise, for high nonwetting saturation only the nonwetting phase flows. 
These two regions are thus effectively single-phase flow in a constraining 
environment consisting of both the solid porous medium and the immobilized 
phase. In-between these regions is the two-phase flow region. The cross-overs 
from single-phase flow to two-phase flow are in fact dynamical phase 
transitions. We wish to study this system with that perspective. As order 
parameter for the transition from single wetting flow to two-phase flow we 
use the nonwetting fractional flow. Similarly we take the wetting fractional 
flow as the order parameter for the transition from single nonwetting flow to 
two-phase flow. In this way the order parameter is zero for single-phase 
flow and nonzero for two-phase flow.

Having defined the order parameter, the nature of possible dynamical phase 
transitions in the system is well illustrated in Fig.\ \ref{fig:fofssample}. 
Starting at the left-hand-side, the transition is from single-phase wetting 
flow to two-phase flow when increasing the nonwetting saturation. Both 
samples show a continuous transition. That is to say that at least to the 
resolution of the curves the order parameter changes continuously from 
zero to nonzero values at the phase boundary. The transition of the curve 
marked with $\circ\ $on the right-hand-side has clear signs of being first 
order. Not 
only can we see how the order parameter is discontinuous, but the one 
isolated data point of $F_{\rm w}=0.0\ $indicates the presence of hysteresis 
or history dependence in the system. Hysteresis is typical for first order 
transitions. In fact, we will use studies of hysteresis in order to unravel 
the nature of the phase diagram. As to the other curve, marked with $\times$, 
on the right-hand-side, the transition seems to be continuous or weakly 
first order.

\subsection{Hysteresis loops}

\begin{figure}
\begin{tabular}{l}
\includegraphics[width=6cm]{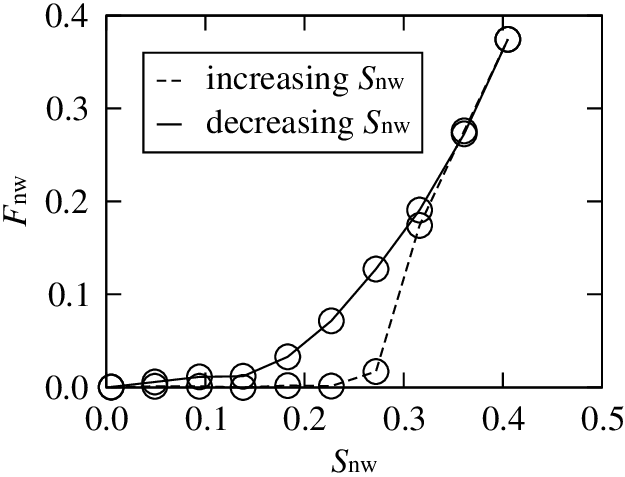} \\
(a) Single-phase wetting flow - two-phase flow \vspace{0.5cm} \\
\includegraphics[width=6cm]{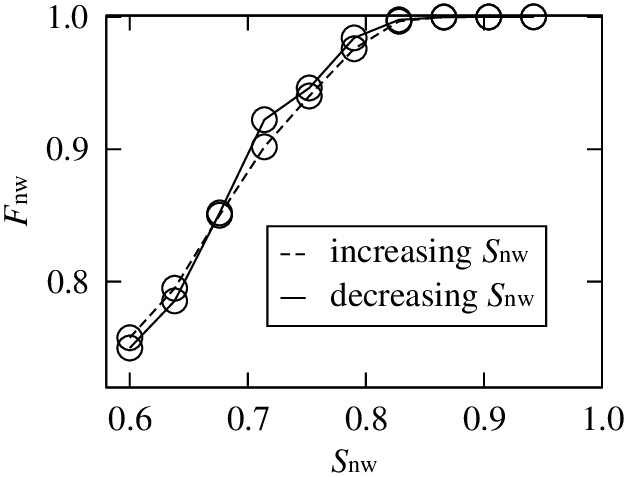} \\
(b) Two-phase flow - single-phase nonwetting flow
\end{tabular}
\caption{The stepwise simulation of two phase transitions. In both cases 
${\rm Ca}=1.00\times10^{-3}$ and $M=10$. The saturation is changed in steps 
up and down past the transition. The system is run for 7 seconds 
(physical time) at each step. (a) Starting at single-phase wetting flow, 
increasing the saturation above the transition to two-phase flow, and then 
lowering the saturation back to the initial level, the system shows 
hysteresis. (b) Starting in the two-phase region and increasing saturation 
to single-phase nonwetting flow, and decreasing again, no hysteresis appears.}
\label{fig:satex}
\end{figure}

In order to map out the phase diagram, we need a tool that can locate the 
phase boundaries in parameter space. The order of the transitions 
should also be determined. Simulations of single points in parameter space 
to steady-state as in Fig.\ \ref{fig:fofssample} are good for the part of 
parameter space where history effects are small. However, whenever one is 
close to a first order transition the history may give data points indicating 
single-phase flow or two-phase flow. It is possible to run simulations 
on several statistically equal realizations of the porous media, as well as 
from various initial configurations, in order to find the region where both 
single-phase and two-phase flow are possible. The computations are somewhat 
computer demanding, and we have found it more efficient to simulate entire 
hysteresis loops to this effect. For instance starting at a specified 
saturation at which the system has two-phase flow, we change in a stepwise 
fashion the saturation slightly, run until steady-state, and measure the 
fractional flow. This is done until the system passes the transition to 
single-phase flow. Thereafter the saturation is changed the opposite way until 
the cross-over from single-phase flow to two-phase flow occurs. In case of 
a continuous transition, these two points of transition will be the same. 
However, in case of a first order transition these points will not be the 
same, and the fractional flow will appear as a hysteresis loop, see 
Fig.\ \ref{fig:satex}(a).

It is also possible to keep the saturation constant and vary some other 
parameter, typically the capillary number. Similar loops are expected 
at first order transitions by this procedure. The advantage of changing 
Ca is that one simply has to change the total flux $Q_{\rm tot}\ $in the 
system. This is a unique operation, while changing the saturation smoothly 
is not unique. It requires that one makes a choice as to where in the system 
the changes should be made. In general, we increase or decrease existing 
bubbles by moving the menisci defining their surface. If the system is very 
fragmented, some smaller bubbles are typically also removed in order to reach 
the desired new saturation.

\subsection{Diagrams for constant Ca}

In this subsection we study one sample of the porous network of size 
$20\times40$. We do not yet go into the possible dependence on size and 
topology, but focus on how the results depend on the three parameters: phase 
saturation, capillary number Ca, and viscosity ratio $M$. First we keep Ca 
and $M\ $constant while we vary the saturation stepwise over the transition 
from single-phase flow to two-phase flow and back.

Examples of transitions between single-phase and two-phase flow are shown in 
Fig.\ \ref{fig:satex}. By varying the saturation both ways over the 
transition, we observe how there can (a), and cannot (b), be hysteresis in 
the transition. Having a transition with little or without hysteresis as in 
Fig.\ \ref{fig:satex}(b), leads to a quite precise determination of the 
transition point. On the other hand it is not so clear where the transition 
occurs for the sample in Fig.\ \ref{fig:satex}(a). Not only are there two 
actual transition points, but these two points should be expected to depend 
on the actual path in parameter space. By that we mean in this case how 
many steps there are in saturation, and for what time the system is 
allowed to relax at each step. Further, the variation from sample to sample, 
or from realization to realization, of systems that are in principle similar, 
may very well be larger when there is hysteresis of this kind. For curves as 
in Fig.\ \ref{fig:satex}(b), the statistical variations are smaller. The 
transitions in Fig.\ \ref{fig:satex} were taken to be at (a) $S_{\rm nw}=0.26$ 
and (b) $S_{\rm nw}=0.84$. Here we employed that convention that the system 
is single-phase whenever it is single-phase in one of the simulated 
directions. This definition is mainly motivated by the need of consistency 
in the extraction of data points in the following.  This is in contrast to
the more conventional definition, where the transition is located at the 
steepest point of the order parameter curve.  However, this definition
leeds to too much noise in the analysis.

The range of capillary numbers studied is 
${\rm Ca}\in(3.16\times 10^{-2},3.16\times 10^{-4})$. On a base 10 
logarithmic scale 9 values of Ca are chosen in steps of $0.25\ $from $-1.50$ 
to $-3.50$. Similarly 21 values of $M\ $are chosen in steps of 
$0.25\ $in the interval ${\rm log}_{10}M\in(-2.50,2.50)$, which 
means that $M\in(-3.16\times 10^{-3},3.16\times 10^{2})$.

\begin{figure*}[t!]
\begin{tabular}{ccccc}
\includegraphics[width=5.5cm]{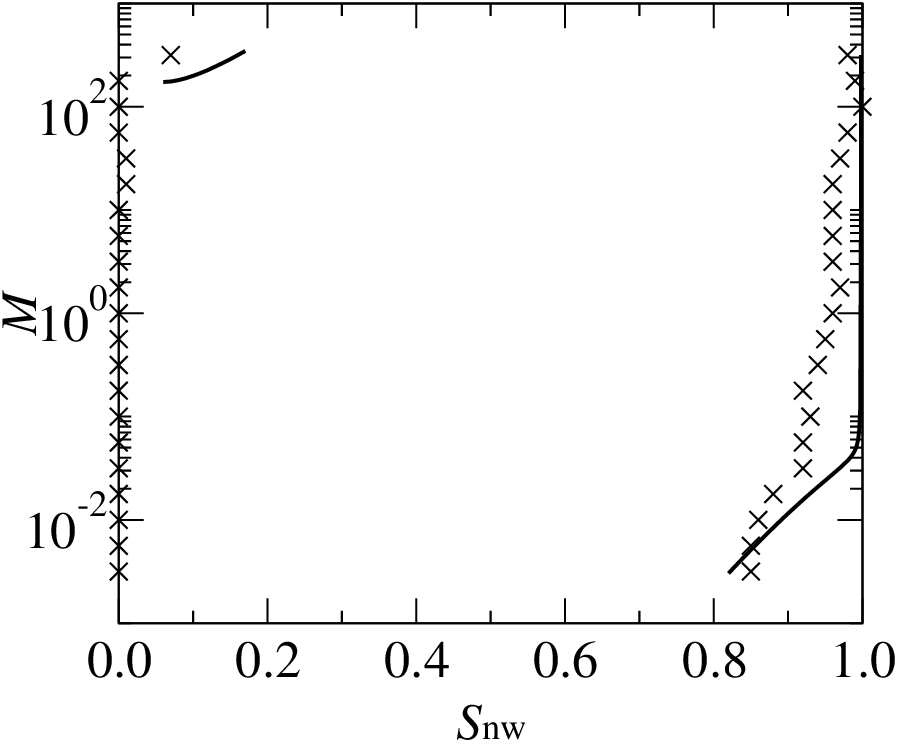}& &
\includegraphics[width=5.5cm]{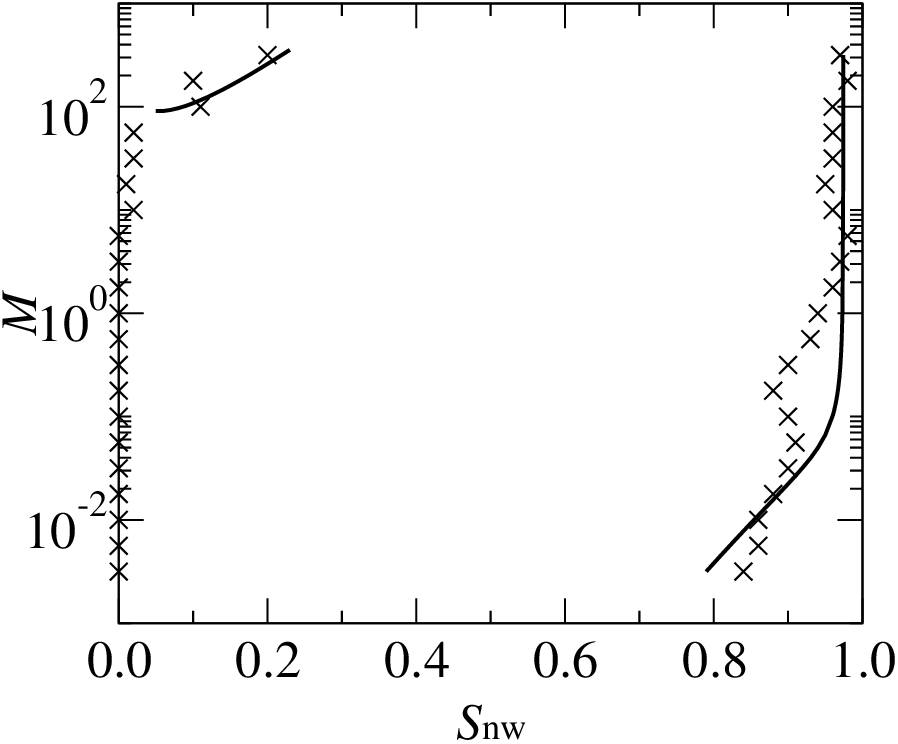}& &
\includegraphics[width=5.5cm]{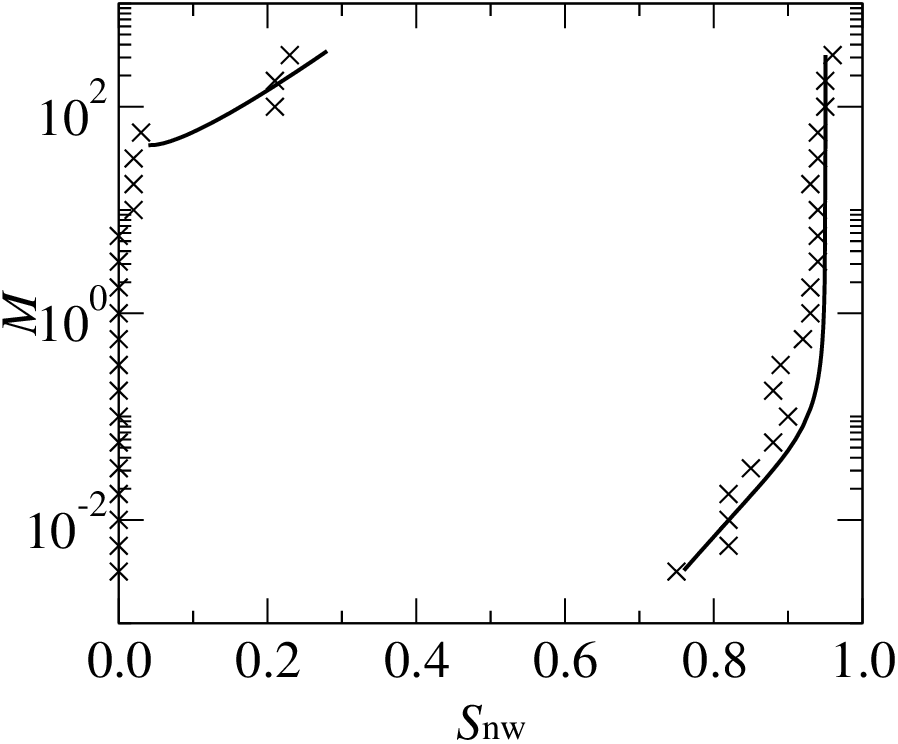} \\
\hspace{0.45cm} (a) ${\rm Ca}=3.16\times 10^{-2}$ & &
\hspace{0.45cm} (b) ${\rm Ca}=1.78\times 10^{-2}$ & &
\hspace{0.45cm} (c) ${\rm Ca}=1.00\times 10^{-2}$ 
\vspace{0.5cm} \\
\includegraphics[width=5.5cm]{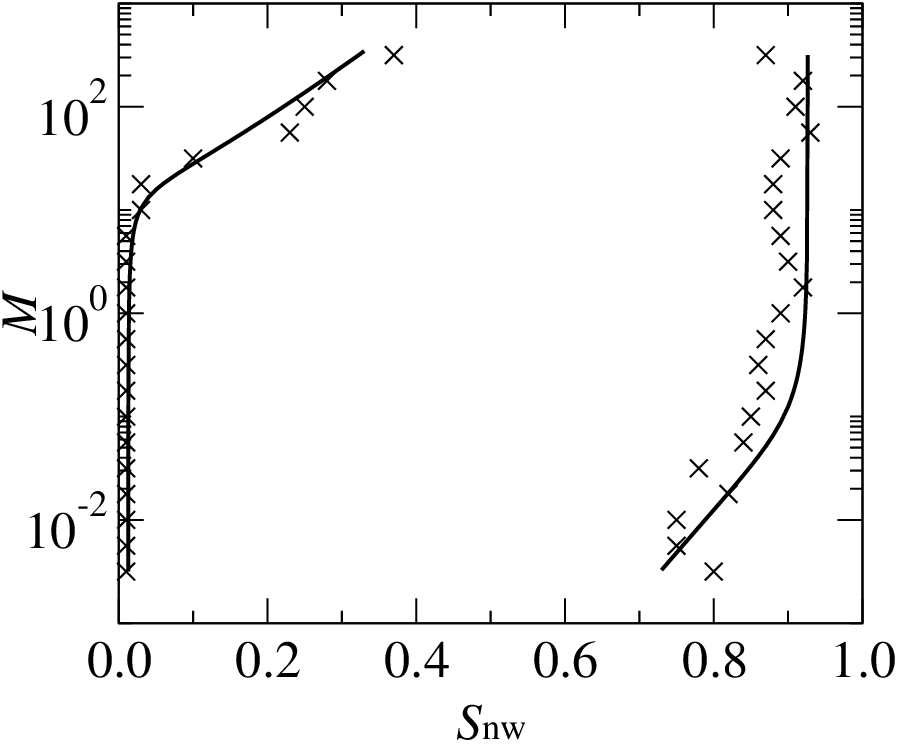}& &
\includegraphics[width=5.5cm]{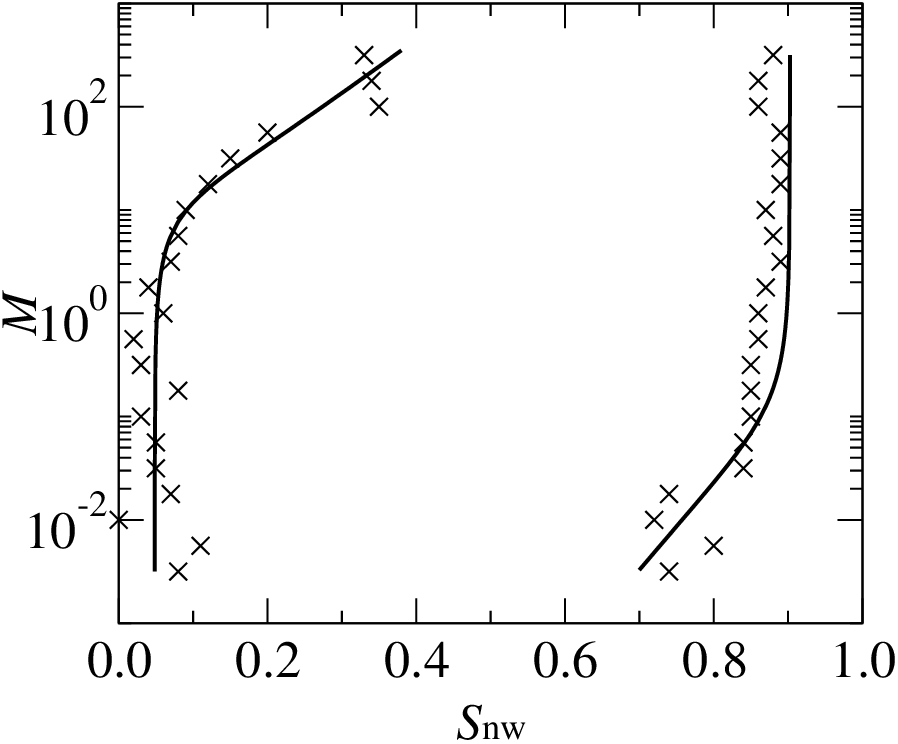}& &
\includegraphics[width=5.5cm]{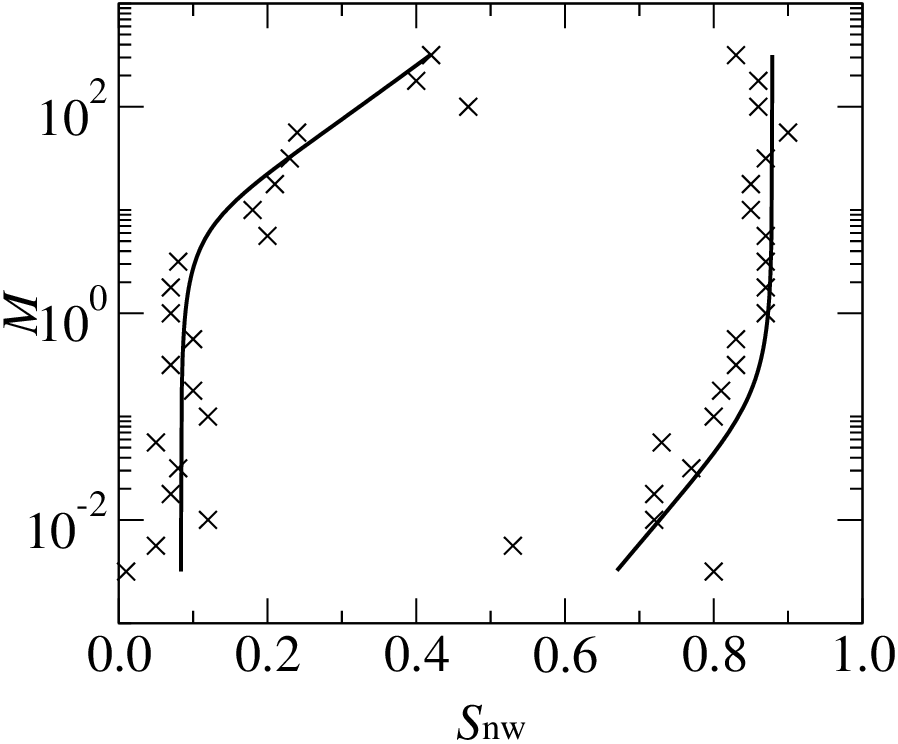} \\
\hspace{0.45cm} (d) ${\rm Ca}=5.62\times 10^{-3}$ & &
\hspace{0.45cm} (e) ${\rm Ca}=3.16\times 10^{-3}$ & &
\hspace{0.45cm} (f) ${\rm Ca}=1.78\times 10^{-3}$ 
\vspace{0.5cm} \\
\includegraphics[width=5.5cm]{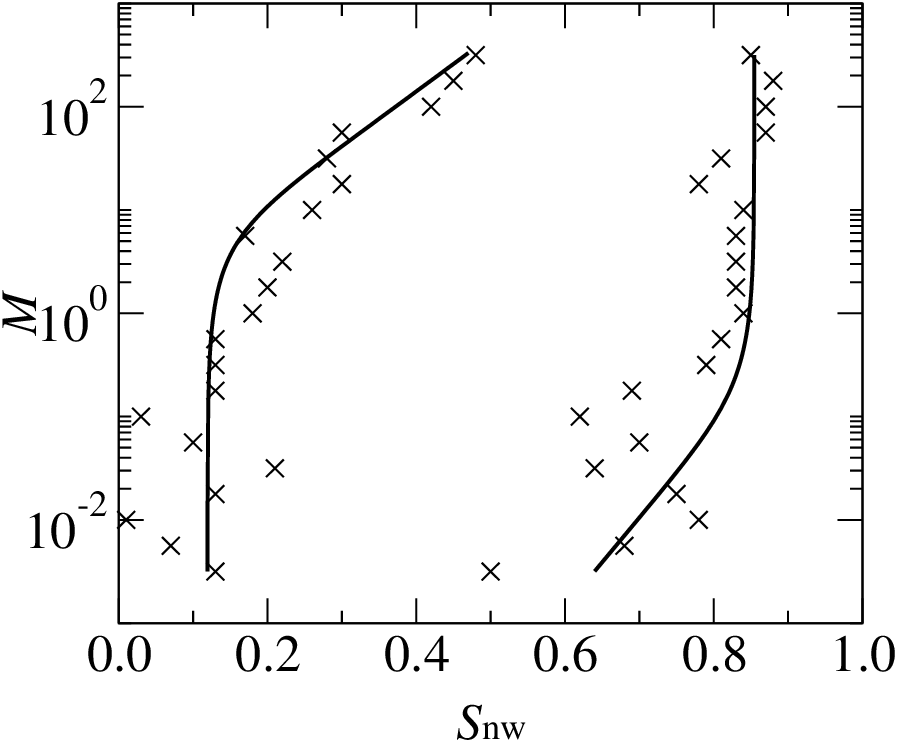}& &
\includegraphics[width=5.5cm]{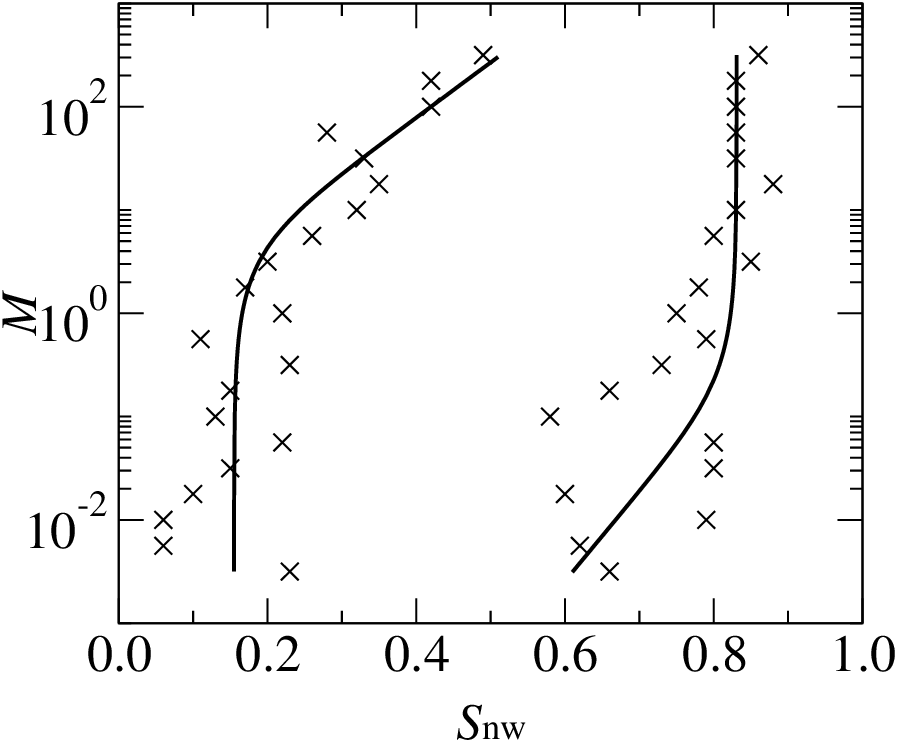}& &
\includegraphics[width=5.5cm]{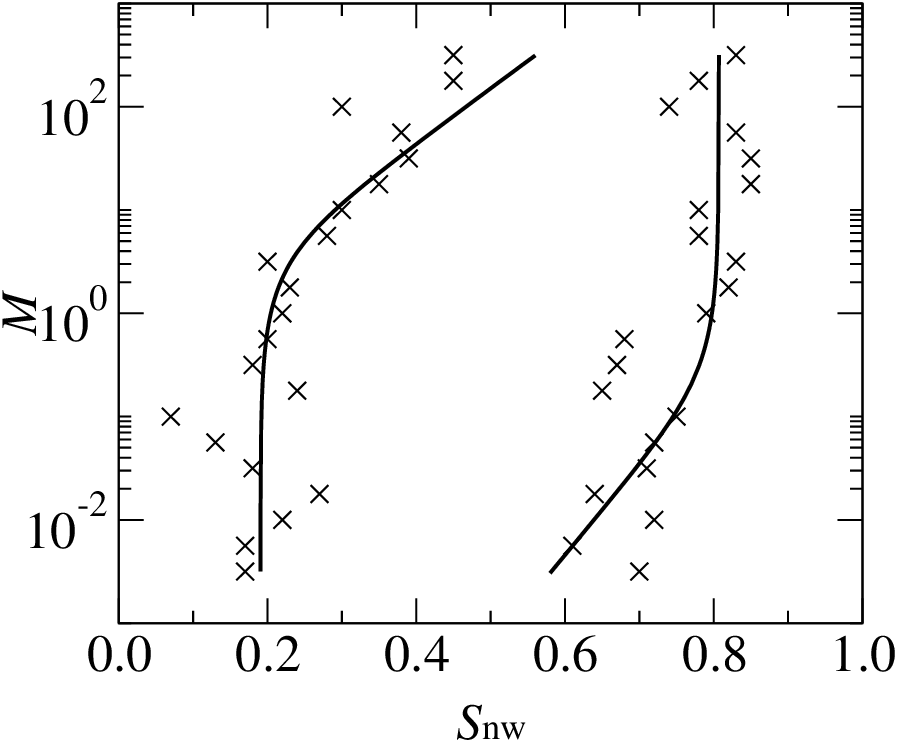} \\
\hspace{0.45cm} (g) ${\rm Ca}=1.00\times 10^{-3}$ & &
\hspace{0.45cm} (h) ${\rm Ca}=5.62\times 10^{-4}$ & &
\hspace{0.45cm} (i) ${\rm Ca}=3.16\times 10^{-4}$ 
\end{tabular}
\caption{The phase diagram for nine values of Ca. The $\times$-marks are the 
transition points from the simulations. The uncertainty in each point is not 
marked, but it is substantial. The scattering of  points in parameter space 
gives an indication of the uncertainty. Phase boundaries are indicated by 
solid lines, see Sec.\ \ref{sec:theory}. The phase diagrams are divided into 
three regions, counting from the left-hand side: single-phase wetting flow, 
two-phase flow, and single-phase nonwetting flow.}
\label{fig:constca}
\end{figure*}

\begin{figure*}[t!]
\begin{tabular}{ccccc}
\includegraphics[width=5.5cm]{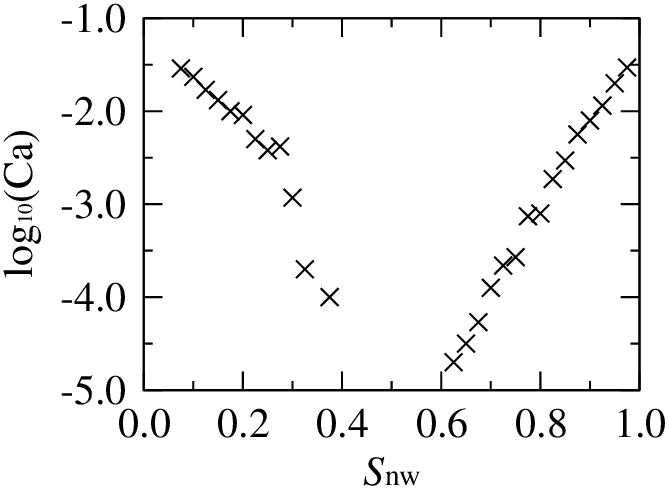}& &
\includegraphics[width=5.5cm]{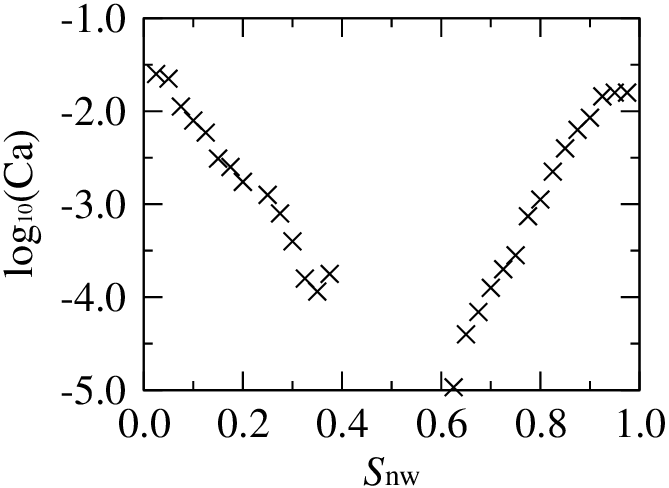}& &
\includegraphics[width=5.5cm]{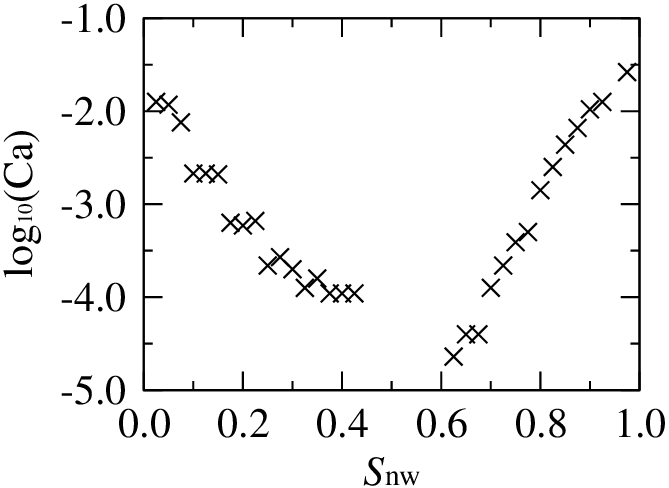} \\
\hspace{0.45cm} (a) $M=10^{2.0}$ & &
\hspace{0.45cm} (b) $M=10^{1.5}$ & &
\hspace{0.45cm} (c) $M=10^{1.0}$ 
\vspace{0.5cm} \\
\includegraphics[width=5.5cm]{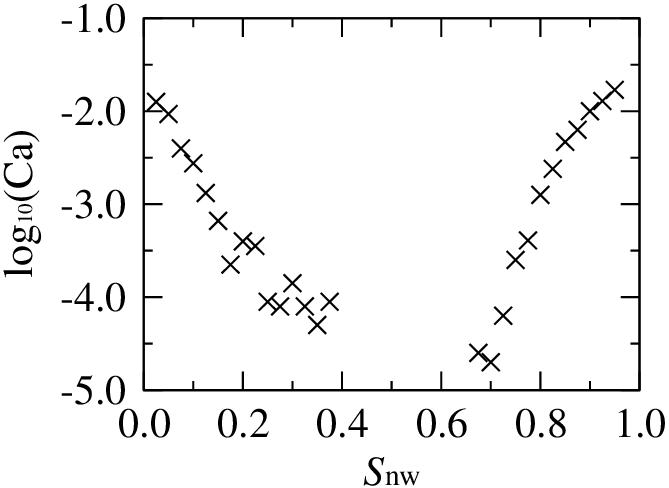}& &
\includegraphics[width=5.5cm]{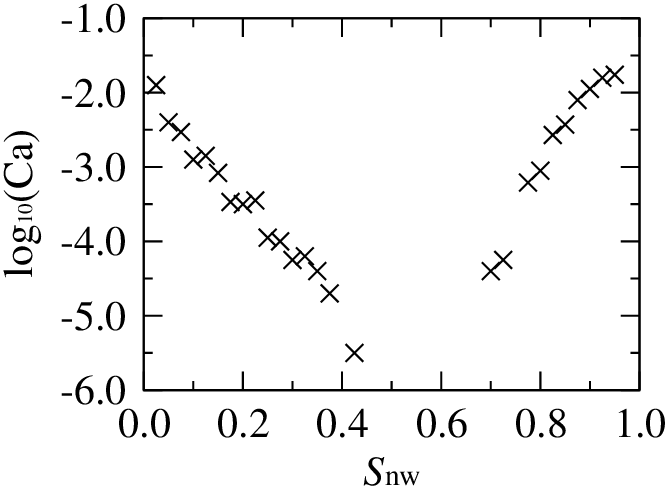}& &
\includegraphics[width=5.5cm]{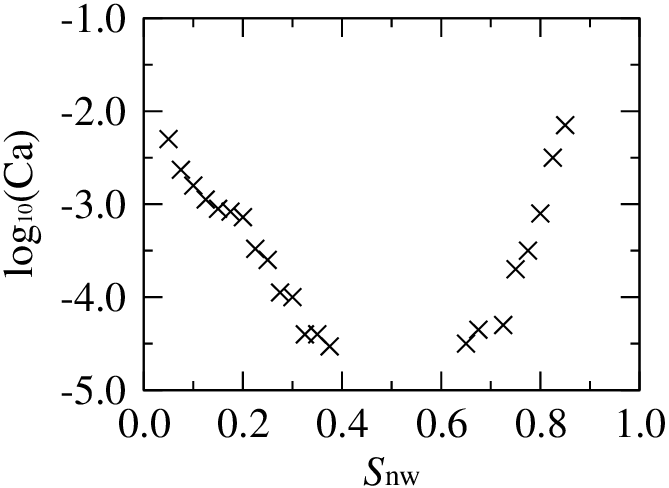} \\
\hspace{0.45cm} (d) $M=10^{0.5}$ & &
\hspace{0.45cm} (e) $M=10^{0.0}$ & &
\hspace{0.45cm} (f) $M=10^{-0.5}$ 
\vspace{0.5cm} \\
\includegraphics[width=5.5cm]{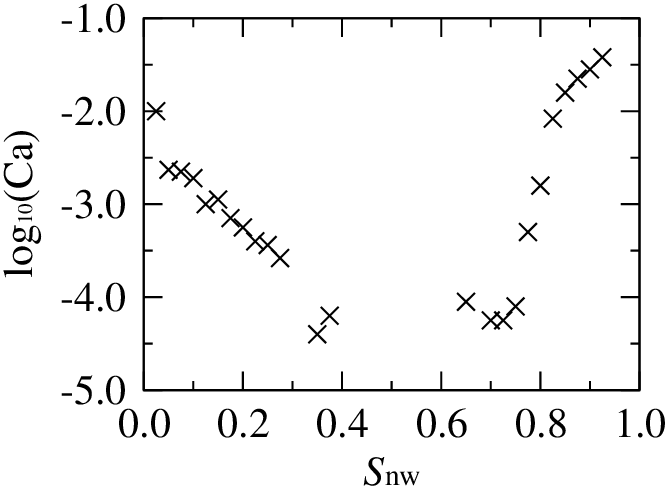}& &
\includegraphics[width=5.5cm]{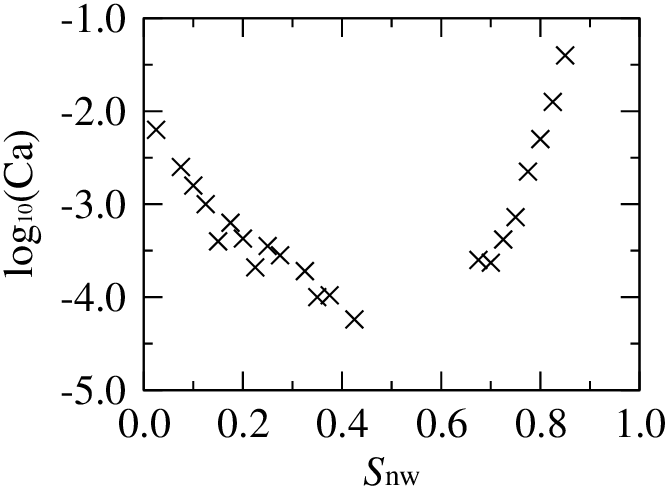}& &
\includegraphics[width=5.5cm]{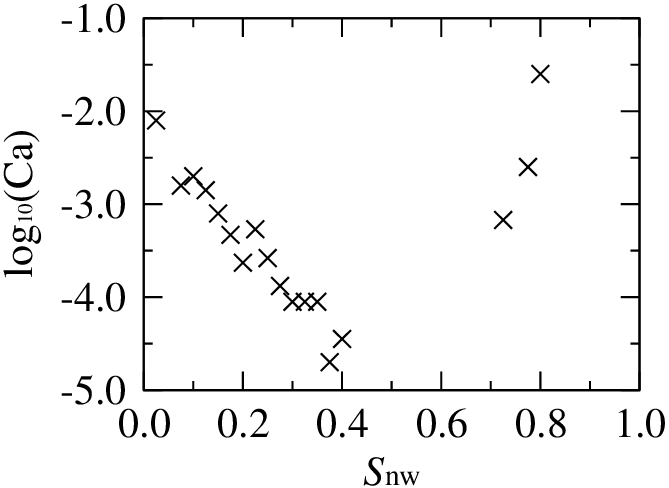} \\
\hspace{0.45cm} (g) $M=10^{-1.0}$ & &
\hspace{0.45cm} (h) $M=10^{-1.5}$ & &
\hspace{0.45cm} (i) $M=10^{-2.0}$ 
\end{tabular}
\caption{The phase diagram for nine values of $M$. The simulations were 
performed at constant saturation, but with varying Ca. The simulated 
points indicate the dynamical phase boundaries. In the lower left part of 
the diagrams, there is single-phase wetting flow; in the middle upper part, 
two-phase flow; and in the lower right part, single-phase nonwetting flow.}
\label{fig:constm}
\end{figure*}

The resulting phase boundaries are shown in Fig.\ \ref{fig:constca} as data 
points. The solid lines are constructed phase boundaries following from 
the semi-empirical theory given in Sec. \ref{sec:theory}. For now, let us 
consider the lines as visual aids to better see the phase boundaries in the 
diagrams. With increasing nonwetting saturation, the two lines separate the 
dynamical phases: single-phase wetting flow, two-phase flow, and single-phase 
nonwetting flow. Observe that for high capillary numbers the two-phase region 
spans almost all of the saturation range. As the capillary number is 
lowered, the single-phase regions grow in size. There is some kind of symmetry 
between the two transitions. That is to say the same effects occur on both 
sides, although not at the same numerical values.

In order to treat the two transitions simultaneously, we employ the following 
notions. Since at each transition there is more volume of the phase that is 
close to single-phase flow, we refer to this phase as the majority phase. 
The other phase is the minority phase. On the left side the majority phase 
is wetting and the minority phase is nonwetting. Regarding viscosity ratios 
$M\ $there are basically three possibilities: viscosity matching phases, 
favorable viscosity ratio, and unfavorable viscosity ratio. Considering the 
system from the point of view of the majority phase, the viscosity ratio is 
favorable when the majority phase is the less viscous phase. Likewise, the 
viscosity ratio is unfavorable when the minority phase is less viscous. On 
the $y$-axes of Fig.\ \ref{fig:constca} $M\ $varies over more than four 
decades. The positive side (on the logarithmic scale) is when the nonwetting 
phase is more viscous than the wetting phase. Thus, the viscosity ratio is 
favorable on the left side and unfavorable on the right side. For negative 
values of ${\rm log}_{10}(M)\ $the ratio is unfavorable on the left side 
and favorable on the right side.

The general structure of the phase diagrams in Fig.\ \ref{fig:constca} is 
as follows. First, consider the case of viscosity matching phases, $M=1$. 
Here, we observe that for high capillary numbers the two-phase region spans 
almost all of the saturation range. As Ca is lowered the single-phase 
regions on each side open up, and the two-phase region becomes smaller. 
We cannot tell from the presented diagrams the limiting behavior of the 
phase boundaries for small Ca. However, since the nine diagrams in Fig.\ 
\ref{fig:constca} are evenly distributed on the log-scale, one can conclude 
that, roughly speaking, the dependence on Ca is logarithmic(also cfr.\ Fig.\ 
\ref{fig:constm}). It is reasonable to expect this behavior to persist also 
for lower Ca.

The dependence on $M\ $is interesting. For all values of Ca the fact that 
each curve has two main parts persists. Roughly, there is one situation for 
favorable viscosity ratios and another for unfavorable viscosity ratios. 
The subsequent discussion of these two situations are made from the two-phase 
flow point of view. The single-phase point of view is taken in 
Sec.\ \ref{sec:theory}.

For unfavorable viscosity ratio (lower left and upper right sides in each of
the diagrams in Fig.\ 
\ref{fig:constca}) the phase boundaries are almost constant with respect to 
the value of $M$. As long as there are clear majority and minority phases, 
and one phase is close to going single-phase, the minority phase consists of 
bubbles and ganglions that are held back by capillary forces. It is always 
the case that the minority phase is held back in this way. However, when 
the viscosity ratio is unfavorable, the minority phase flows more easily. 
The viscous drag on the minority phase is thus negligible for this range of 
$M$-values explaining the constant location of the boundaries. This can be 
understood so that when the minority phase flows sufficiently easy, then it 
simply follows the majority phase around when capillary forces are overcome. 
Notice that the noise level increases with decreasing Ca. When the capillary 
forces become relatively stronger the motion of bubbles involves larger 
fluctuations, so this behavior is expected when properties are averaged 
over the same amount of time.

A different situation appears for favorable viscosity ratios. The capillary 
forces and viscous forces both try to hold back the minority phase. In this 
situation the actual viscosity ratio plays a role. The more favorable 
viscosity ratio, the more will the minority phase slow down. Because of 
the capillary forces the minority phase will not only slow down, but 
actually stop, and single-phase flow results. The border between two-phase 
flow and single-phase flow varies with  $M$, as can be seen in Fig.\ 
\ref{fig:constca}. We observe the general trend that the single-phase region 
increases logarithmically with increasingly favorable viscosity ratio.

%The positioning of the cross-over points has to do with the competition 
%between capillary and viscous forces. It is a function of Ca, but 
%maybe not in the way that one should expect. As the capillary number 
%is lowered, meaning increasingly dominant capillary forces, the 
%single-phase regions expand. This is reasonable since the capillary 
%forces can then sustain stronger viscous pressure and thus hold back 
%more of the minority phase. One could expect that the $M$-value of the 
%cross-over point would move so that these constant region would increase 
%with lower Ca. The reason would have been that when the capillary 
%forces are stronger, they would dominate over more favourable viscosity 
%ratios. As we can observe, the effect is quite the opposite, which 
%means that this reasoning is wrong. Instead there are inter-play 
%and co-operation between viscous and capillary forces. The effect 
%is additive in the sense that when the viscous forces are stronger, 
%less help from favourable viscosity ratio is necessary to reach 
%the cross-over point.

\subsection{Diagrams for constant $M$}

So far we have discussed the results from simulations where the saturation 
has been varied over the dynamical phase transition. In this subsection 
we present results from simulations in which the saturation and the 
viscosity ratio are held constant, while Ca is varied down and up 
again over the transition. Because of history effects and the fact that 
saturation and Ca are not varied in the same way, it is a priori not 
clear that the phase boundaries will be the same. Therefore, we check 
the results for consistency.

For nine different values of $M$, equidistantly distributed on a logarithmic 
scale between $M=1/100\ $to $M=100$, simulations have been performed for a 
large set of saturations. The results are shown in Fig.\ \ref{fig:constm}. 
For large viscosity contrasts it turned out to be difficult to simulate very 
small saturations of the minority phase, possibly because the transition 
would occur for very large capillary numbers. More importantly, the central 
part of the saturation range was also difficult to simulate, because the 
fluctuations in the systems are larger in this region. One might expect that 
it is possible for the system to reach either single-phase wetting flow or 
single-phase nonwetting flow simply as the result of fluctuating into that 
state. We found that the fluctuations remain large for capillary numbers 
all the way down to ${\rm Ca} = 10^{-6}\ $for several samples. It is also 
possible that there is a finite saturation range, where the dynamical phase 
is two-phase flow for all values of Ca. We conclude that this region 
therefore may be out of reach. Even if we could get the results from the 
simulator, such low capillary numbers are outside the range where this 
simulator works well, and the relevance to physical systems would be limited. 

The observations we can make from Fig.\ \ref{fig:constm} are as follows. 
The nine presented diagrams have basically the same appearance. They 
consist of three parts: single-phase wetting flow in the lower left corner, 
single-phase nonwetting flow in the lower right corner, and two-phase 
flow in the upper middle region.

\begin{figure}
\includegraphics[width=6cm]{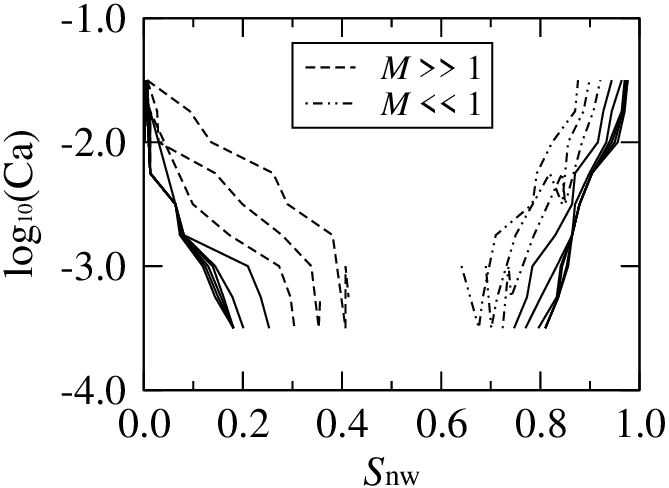} \\
(a) Extracted from Fig.\ref{fig:constca} \vspace{0.5cm} \\
\includegraphics[width=6cm]{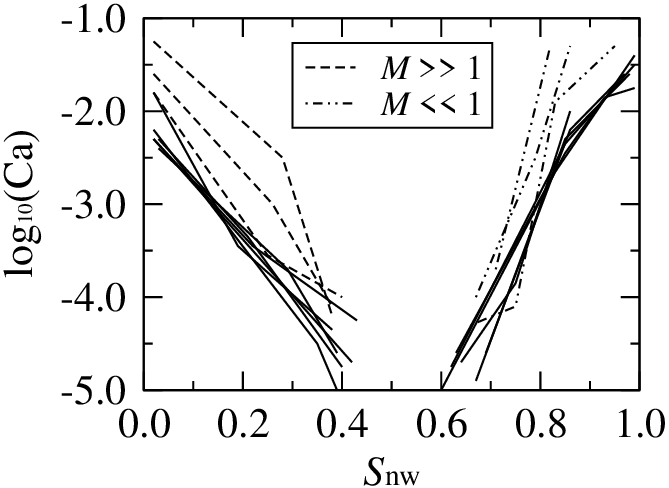} \\
(b) Extracted from Fig.\ref{fig:constm}
\caption{The phase boundaries from Figs.\ \ref{fig:constca} and 
\ref{fig:constm} are extracted as lines in a $S_{\rm nw}-{\rm Ca}$ 
coordinate system. Here, $M\ $parametrizes the curves and takes the 
same nine values as in Fig.\ref{fig:constm}.}
\label{fig:constmall}
\end{figure}

First, observe that as far as simulations have been performed, the phase 
boundaries have ending points at zero and unity saturation for finite values 
of Ca. At higher values of Ca the system has two-phase flow for all 
saturations. This is not surprising. High capillary numbers mean vanishing 
small capillary forces relative to viscous forces. In principle, as long 
as the viscosity contrast is finite or not too large, the more viscous phase 
would still flow unless there also are capillary forces to keep it from 
flowing. Hence, for these viscosity ratios, the capillary forces are too 
small above some threshold to prevent one phase from flowing.

Second, the general behavior of the two boundaries away from their end 
points, is that as Ca is lowered, the minority saturation increases. 
The meaning is simply that when the capillary forces become relatively 
stronger the single-phase regions grow in size. For low Ca values the 
two-phase region becomes quite narrow. As we already have pointed out, 
we cannot tell whether the two boundaries meet in a triple point or if 
there will be a finite two-phase region also for smaller Ca.

In order to compare the curves in Fig.\ \ref{fig:constm}, the results have 
been extracted and put into one diagram, see Fig.\ \ref{fig:constmall}(b). 
The curves overlap considerably so we have chosen to use the same line 
style for most of the lines. The exception is that for the three most 
favorable viscosity ratios of each transition a different line style is used. 
The three curves separate more and more from the rest of the curves 
with increasing viscosity contrast.

To compare the simulation results shown in Figs.\ \ref{fig:constca} and 
\ref{fig:constm}, we exhibit data from Fig.\ \ref{fig:constca} in Fig.\ 
\ref{fig:constmall}(a). Here, the same set of viscosity ratios were used 
as in Fig.\ \ref{fig:constmall}(b). This was done for best possible 
comparison. The line styles are also the same. Note that the range of 
capillary numbers in Fig.\ \ref{fig:constmall}(a) is smaller than in Fig.\ 
\ref{fig:constmall}(b). The results are not identical, but nevertheless 
quite close to each other within the overlapping range. We conclude 
that the two ways of performing the simulations give consistent results.

There are some general features of the diagrams in Fig.\ \ref{fig:constmall}. 
First we observe that for both the left side boundary and the right side 
boundary, there seems to be a lower limit. This means that for all 
viscosity ratios, the respective single-phase region on each side extends 
at least to this limit. Roughly speaking these limits vary logarithmically 
with Ca. Again we cannot tell if this behavior will continue for lower 
Ca than depicted.

The other feature is that for higher Ca, say ${\rm Ca} > 10^{-3.5}$, 
favorable viscosity ratio affects the phase boundary. To be precise, for 
the three intermediate values of $M$: $M=10^{-0.5}$, $M=1$, and $M=10^{0.5}$, 
the phase boundaries on both sides are more or less equal to the 
limiting behavior just discussed. For these values of $M$, 
$M=10^{1.0}$, $M=10^{1.5}$, and $M=10^{2.0}$, the viscosity ratio is 
favorable for the transition on the left side between single-phase wetting 
flow and two-phase flow. The corresponding three lines are marked with dashed 
lines. These values of $M\ $are unfavorable for the right-hand side 
transition. The last three values of $M$, $M=10^{-1.0}$, $M=10^{-1.5}$, 
and $M=10^{-2.0}$, are favorable for the right-hand side transitions. They 
are indicated with dash-dotted lines. For the left-hand side these $M\ $values 
are unfavorable. In conclusion, there are six lines that are singled 
out for their favorable viscosity ratio. They clearly show how the 
single-phase regions become larger as the viscosity ratio becomes 
increasingly favorable. This effect can be seen in both Figs.\ 
\ref{fig:constmall}(a) and \ref{fig:constmall}(b).

\section{Location of the phase boundaries}
\label{sec:theory}

In the previous subsection the point of view taken was that of the two-phase 
flow system. That is to say, we used the dynamical definition of the 
capillary number, Eq.\ (\ref{eq:cadef}), and we asked when does one of 
the phases cease to flow. Reversing the point of view is fruitful. 
Consider the system starting from a single-phase flow situation, 
the natural question is when does the second phase start to flow.

Being in a single-phase flow situation, although with both phases present in 
the system, means that the capillary number should be defined differently. 
Namely, the relevant viscosity is no longer the volume averaged viscosity, 
but the viscosity of the fluid that is actually flowing. The wetting and 
nonwetting capillary number is defined as
\begin{equation}
\label{eq:cawdef}
  {\rm Ca_w} = \frac{Q_{\rm tot}\mu_{\rm w}}{\Sigma\gamma}\ ,
\end{equation}
and
\begin{equation}
\label{eq:canwdef}
  {\rm Ca_{nw}} = \frac{Q_{\rm tot}\mu_{\rm nw}}{\Sigma\gamma}\ ,
\end{equation}
respectively. Here, ${\rm Ca_w}$ is the relevant capillary number for 
single-phase wetting flow, whereas, ${\rm Ca_{nw}}$ is relevant for 
single-phase nonwetting flow. 

As opposed to the qualitative statements regarding the phase boundaries 
that were made in Sec.\ \ref{sec:diagram}, these definitions of the 
capillary number are also quantitatively useful. For a given system, i.e.\ 
given topology and saturation, the transition to two-phase flow takes 
place at a given wetting (or nonwetting) capillary number. That means 
with varying value of the viscosity ratio $M$, which also changes 
${\rm Ca_{dyn}}$, the relevant single-phase capillary number still 
holds a fixed value at the transition. This is plausible as long as 
the transition to two-phase flow is continuous. In that case the amount 
of minority phase flowing just after its mobilization is still only 
a small fraction of the total flow, and therefore the physical 
effective viscosity is still very close to the viscosity of the 
single-phase that just before flew alone. However, should the transition 
be first order so that a finite fraction is mobilized at once, 
then for large variations in $M$ some corrections should be made. 
Nevertheless, for now we assume that the single-phase capillary number 
governs the location of the phase boundaries. The question of the order 
of the transition is addressed in Sec.\ \ref{sec:ordertrans}, and 
thereby the validity of this assumption is discussed.

\begin{figure}[h!]
\includegraphics[width=6cm]{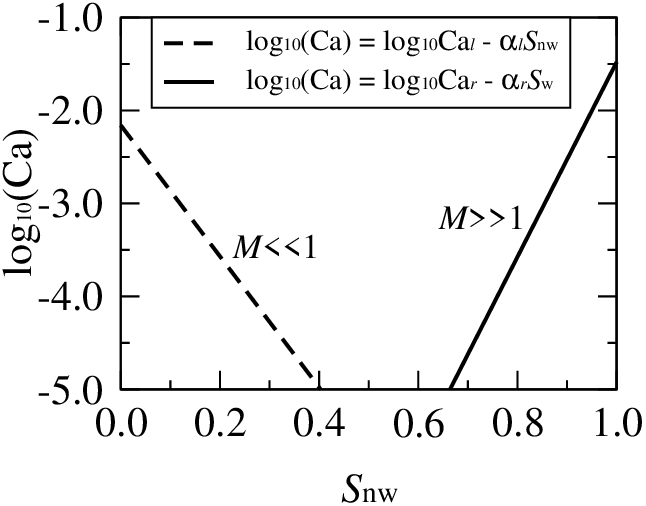}\\
\caption{The phase boundaries in Fig.\ \ref{fig:constmall} approach 
a limiting boundary for low $M$ (left side) and large $M$ (right side), 
and these limits are shown here. The fitted values for the constants 
are: ${\rm Ca}_l=10^{-2.16}$, ${\rm Ca}_r=10^{-1.48}$, $\alpha_l=7.04$, 
and $\alpha_r=10.42$.
}
\label{fig:Mline}
\end{figure}

We can write the dynamical capillary number
\begin{equation}
\label{eq:cadynlong}
  {\rm Ca_{dyn}} = \frac{Q_{\rm tot}\mu_{\rm w}}{\Sigma\gamma}
  \left( S_{\rm w} + MS_{\rm nw} \right)\ ,
\end{equation}
from which it follows that in the limit of small $M$
\begin{equation}
\label{eq:CalimMl}
  {\rm Ca_{dyn}} = {\rm Ca_{w}}S_{\rm w}\ ,
\end{equation}
and in the limit of large $M$
\begin{equation}
\label{eq:CalimMr}
  {\rm Ca_{dyn}} = {\rm Ca_{nw}}S_{\rm nw}\ .
\end{equation}
In Fig.\ \ref{fig:constmall} we observe the limiting behavior for 
unfavorable $M$ of the phase boundaries. There is on each side a 
limiting line, below which there is always single-phase flow. By Eqs.\ 
(\ref{eq:CalimMl}) and (\ref{eq:CalimMr}) these limits are identified 
with the certain wetting or nonwetting capillary number that for each 
value of the saturation determines the location of the transition. The 
limiting lines and the fitted functional relationships are shown in Fig.\ 
\ref{fig:Mline}. We take this logarithmic dependency as an empirical 
input, and see which predictions can be made therefrom employing theory.

We want to calculate the location of the phase boundaries for fixed dynamical 
capillary numbers as shown in Fig.\ \ref{fig:constca}. Considering first 
the left hand side (transition from wetting single-phase to two-phase flow), 
the criterion for the phase boundary is expressed as
\begin{equation}
\label{eq:criterion}
  {\rm Ca_w}S_{\rm w} = {\rm Ca}_l 10^{-\alpha_l S_{\rm nw}}\ ,
\end{equation}
thus being a function of saturation. Letting ${\rm Ca_{dyn}}$ be fixed, 
this criterion inserted into Eq.\ (\ref{eq:cadynlong}) gives when solving 
for $M$ as a function of $S_{\rm nw}$
\begin{equation}
\label{eq:soll}
  M = \frac{1-S_{\rm nw}}{S_{\rm nw}} \left(
      \frac{{\rm Ca_{dyn}} 10^{\alpha_l S_{\rm nw}}}{{\rm Ca}_l} - 1
      \right)\ .
\end{equation}
The interpretation of Eq.\ (\ref{eq:soll}) is that for a fixed value of 
${\rm Ca_{dyn}}$ the phase boundary is given by the set of values of 
$M$ and $S_{\rm nw}$ that fulfill this relationship. For values of 
${\rm Ca_{dyn}}>{\rm Ca}_l$, the wetting single-phase region does not exist, 
and so does the identification in the limit, Eq.\ (\ref{eq:CalimMl}) not 
apply. Still the criterion Eq.\ (\ref{eq:criterion}) can be used, and the
result is that the single-phase region can only be reached for values of $M$ 
larger than a certain threshold, see Fig.\ \ref{fig:constca}(a)-(c). This is 
not the case for ${\rm Ca_{dyn}}<{\rm Ca}_l$, where there is a single-phase 
region for all $M$.

In a similar way the boundaries on the right hand side between two-phase 
flow and single-phase nonwetting flow is found to obey
\begin{equation}
\label{eq:solr}
  M = \left[ \frac{1-S_{\rm w}}{S_{\rm w}} \left(
      \frac{{\rm Ca_{dyn}} 10^{\alpha_r S_{\rm w}}}{{\rm Ca}_r} - 1
      \right)\right]^{-1}\ .
\end{equation}
By using the Eqs.\ (\ref{eq:soll}) and (\ref{eq:solr}) the solid lines in 
Fig.\ \ref{fig:constca} are the constructed phase boundaries for the 
respective values of ${\rm Ca_{dyn}}$. The general good agreement between 
the simulated data points and these semi-empirical, semi-theoretical 
lines, serves as a confirmation a posteriori of this theoretical approach. 
Indeed, we conclude that the single-phase capillary number is the 
relevant entity that controls the transition to two-phase flow, rather than
the more commonly used capillary number defined in Eq.\ (\ref{eq:cadef}).
However, we should point out that from an experimental point of view, it is
the conventionally defined capillary number which is the most relevant one.

\section{Order of the transitions}
\label{sec:ordertrans}

\begin{figure}[t!]
\includegraphics[width=6cm]{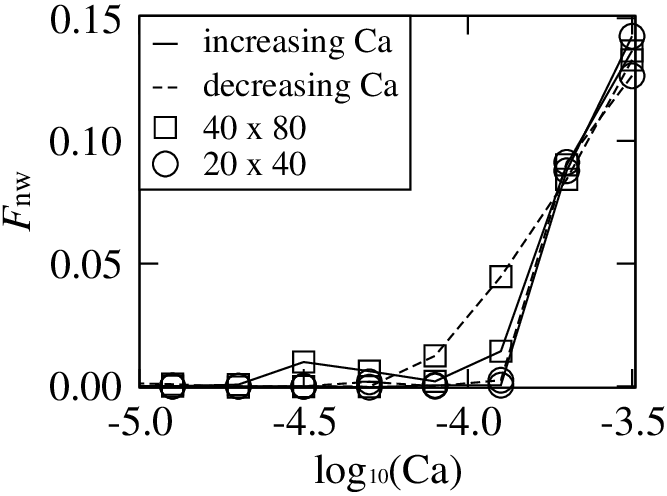}\\
(a) $M=1$
\vspace{0.5cm}\\
\includegraphics[width=6cm]{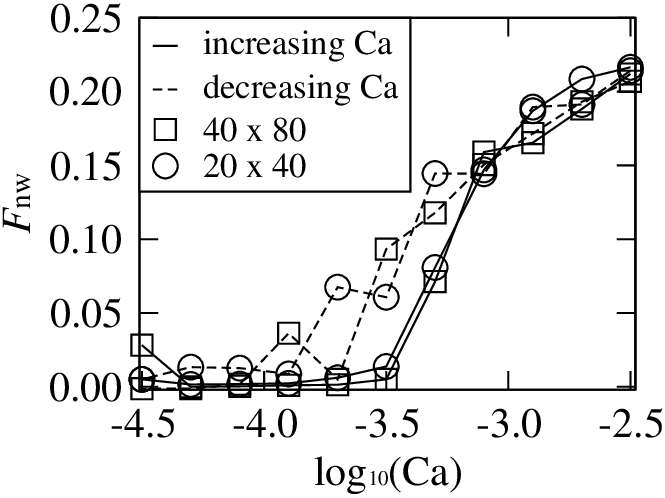}\\
(b) $M=10$
\caption{Two samples of simulations on two different system sizes. The 
saturation is $S_{\rm nw}=0.3$ in both cases. The value of $M$ is the only 
difference between (a) and (b). The transitions are (a) continuous or 
weakly first order, and (b) first order. The important aspect is that the 
hysteresis effects do not get smaller with increasing system size.}
\label{fig:ssloop1}
\end{figure}

So far we have only discussed the location of the dynamical phase boundaries. 
The order of the transitions across these boundaries should also be 
discussed. All the simulations presented so far were done on a single 
system size, namely $20\times 40\ $nodes. It is necessary to take into 
account effects of changing the system size to determine the order 
of the transitions. However, first we present results for this system size.

Each simulation in Fig.\ \ref{fig:constca} can be categorized into 
showing or not showing hysteresis. We look for the general trend while 
accepting that fluctuations and randomness play a role. Some realizations 
do not show hysteresis even though all neighboring points do, and vice 
versa. It turns out that the phase boundaries of Fig.\ \ref{fig:constca} 
can be divided into two logical parts. The almost vertical line segments 
where the systems have unfavorable viscosity ratio differ from the 
other parts of the curves where the viscosity ratio is favorable. In 
the vertical parts the simulations show no or little signs of hysteresis 
in the order parameter. This indicates that the transitions in these 
regions are continuous. The fact that the curves are noisy for lower Ca 
in these regions, does not affect the order. In the sloping parts of the 
boundaries the order parameter exhibit discontinuity and hysteresis in 
most simulations. The level of hysteresis increase with increasing 
viscosity contrast. We claim that the transitions in these regions 
are first order.

Whether a transition is continuous or first order is a property connected 
to a specific path in parameter space. However, the paths in Fig.\ 
\ref{fig:constca} and in Fig.\ \ref{fig:constm} all mean a transition from 
single-phase flow to two-phase flow. We find similar hysteresis effect 
for both kinds of paths through parameter space.

In general, it is possible that the appearance of different orders of the 
transitions is a result of finite size effects. We have performed some 
selected simulations on a larger system size; $40\times 80\ $nodes. 
Comparisons between the results of the two system sizes are found in 
Fig.\ \ref{fig:ssloop1}. The first order transition in (b) persist 
also for the larger system. If the signs of first order had been weaker 
for larger system sizes, then finite size effects had been the cause for 
apparent first order transitions. However, this is not the case here.

The noisy character of the vertical parts, showing continuous transitions, 
can be attributed to the geometrical heterogeneity of the system. 
Some parts of the system are more active in transportation than others. 
In such rather small systems self-averaging is small and by chance the 
minority phase is to a larger or smaller extent placed in inactive parts 
of the system. This gives a shift in position of the transition, i.e.\ noise, 
but does not affect the correlation between minority phase bubbles that 
are placed in active regions of the system. In fact these bubbles that are 
only held back by capillary forces are all correlated upon onset of 
mobilization. The motion of one bubble in one place relieves the pressure 
on all other minority bubbles in the system. The opposite is also true, 
when this bubble stops again, the pressure on the others increase immediately 
and this is likely to cause the onset of mobilization elsewhere. This is 
consistent with having infinite correlation length in statistical 
mechanics at continuous transitions.

The sloped sections of the boundaries, which from the hysteresis loop 
consideration show signs of first order transitions, show different 
characteristics. The viscosity ratio is unfavorable and the mobilization 
requires more pressure on a bubble, but when mobilized, larger bubbles or 
connected bubbles are involved. However, this effect is more local in the 
sense that the minority phase is not moving a little bit here and a little 
bit there in the form of different bubbles, but rather one larger region
moves for a longer time when it is first mobilized. This correspond to a 
non-infinite correlation length. Note that the system is driven by the 
global criterion of fixed total flux through the system. From this point 
of view, everything is all the time correlated with everything, so when 
we speak about correlations, we speak about the motions of the minority 
phase as they are observed.

\section{Conclusion}

This study concerns steady-state properties of two-phase flow in porous 
media. Average flow properties are monitored by using a network simulator 
on pore level. A systematic change of the system parameters: phase 
saturation, viscosity ratio, and capillary number, is performed. We 
demonstrate how the system can be in either one of three dynamical states 
or phases. These are: single-phase wetting flow, two-phase flow, and 
single-phase nonwetting flow. Upon passing from one dynamical phase to 
another the system undergoes a dynamical phase transition. The phase 
diagram for the dynamical phases is revealed and its properties are discussed.

By means of hysteresis we determine the order of the phase transitions 
and we find that both continuous and first order transitions occur 
across different parts of the phase-boundaries. This is connected to 
two very different situations. One situation is when the viscosity ratio 
is defined to be favorable with respect to the majority phase: both 
capillary forces and viscous forces hold back the minority phase from 
flowing, next to this transition. The other case is when the viscosity 
ratio is defined to be unfavorable: only capillary forces try to hold 
back the minority phase next to the transition.

The phase diagram that is provided is for steady-state properties. 
This should be set in perspective to the phase diagrams by Lenormand 
\emph{et al.} for drainage invasion flow properties\cite{LZS83,LTZ88}. 
Although the fact that the system investigated is a particular 
steady-state two-phase flow system, we argue that the general 
structure of the dynamical phase diagram has a universally valid structure.

We find that the definition of the capillary number is of major 
importance. Whereas for the actual two-phase flow region we employ a 
volume averaged viscosity in the definition, it turns out that the 
actual locations of the phase boundaries are determined by the capillary 
number based on the viscosity of the single phase in question. Based on 
the relationships between the capillary numbers, theoretical part, 
and limiting  of the phase boundaries, found by the simulations, 
we establish a semi-empirical theory for the location of the phase boundaries. 
Upon direct comparison this theory is found to be in agreement with the 
simulated data points.

\begin{acknowledgments}
H.A.K.\  thanks VISTA, a collaboration between Statoil and the Norwegian
Academy of Science and Letters, for financial support.  We thank E.\ Skjetne
for very valuable discussions, in particular in connection with 
Section \ref{sec:theory} where his contribution was substantial.
\end{acknowledgments}

\bibliography{referanser}

\end{document}